\documentclass[aps,prl,twocolumn,showpacs,superscriptaddress,groupedaddress]{revtex4}  
\usepackage{amsmath}
\usepackage{amssymb}
\usepackage{amstext}
\usepackage{amsopn}
\usepackage{amsfonts}
\usepackage{amsxtra}
\usepackage[english]{babel}
\usepackage{graphicx}
\usepackage{bm}
\usepackage{multirow}
\usepackage{dcolumn}
\usepackage{color}
\usepackage{hyperref}
\usepackage{todonotes}
\usepackage{verbatim}
\usepackage{textcomp}
\usepackage{soul}

\begin{document}

\title{Anisotropic Correlated Electronic Structure of Colossal Thermopower Marcasite FeSb$_2$}

\author{A. Chikina}
\thanks{these authors contributed equally}
 \address{Swiss Light Source, Paul Scherrer Institut, CH-5232 Villigen, Switzerland.}
\author{J.-Z. Ma}
\thanks{these authors contributed equally}
 \address{Swiss Light Source, Paul Scherrer Institut, CH-5232 Villigen, Switzerland.}
\author{W. H. Brito}
\thanks{these authors contributed equally}
 \address{Condensed Matter Physics and Materials Science Department, Brookhaven National Laboratory, Upton 11973 New York USA.}\address{Department of Physics and Astronomy, Rutgers, The State University of New Jersey - Piscataway, NJ 08854, USA.}
\author{S. Choi}
 \address{Condensed Matter Physics and Materials Science Department, Brookhaven National Laboratory, Upton 11973 New York USA.}
\author{Q. Du}
 \address{Condensed Matter Physics and Materials Science Department, Brookhaven National Laboratory, Upton 11973 New York USA.}\address{School of Materials Science and Engineering, Stony Brook University, Stony Brook 11790 New York USA.}
\author{J. Jandke}
 \address{Swiss Light Source, Paul Scherrer Institut, CH-5232 Villigen, Switzerland.}
\author{H. Liu}
 \address{Swiss Light Source, Paul Scherrer Institut, CH-5232 Villigen, Switzerland.}
\author{N. C. Plumb}
 \address{Swiss Light Source, Paul Scherrer Institut, CH-5232 Villigen, Switzerland.}
\author{M. Shi}
 \address{Swiss Light Source, Paul Scherrer Institut, CH-5232 Villigen, Switzerland.}
\author{C. Petrovic}
 \address{Condensed Matter Physics and Materials Science Department, Brookhaven National Laboratory, Upton 11973 New York USA.}\address{School of Materials Science and Engineering, Stony Brook University, Stony Brook 11790 New York USA.}
\author{M. Radovic}
\email{milan.radovic@psi.ch}
 \address{Swiss Light Source, Paul Scherrer Institut, CH-5232 Villigen, Switzerland.}
\author{G. Kotliar}
\email{kotliar@physics.rutgers.edu}
\address{Condensed Matter Physics and Materials Science Department, Brookhaven National Laboratory, Upton 11973 New York USA.}\address{Department of Physics and Astronomy, Rutgers, The State University of New Jersey - Piscataway, NJ 08854, USA.}
\date{\today}

\begin{abstract}
Iron antimonide (FeSb$_2$) is a mysterious material with peculiar colossal thermopower of about $-45$ mV/K at 10 K. However, a unified microscopic description of this phenomenon is far from being achieved. The understanding of the electronic structure in details is crucial in identifying the microscopic mechanism of FeSb$_2$ thermopower. Combining angle-resolved photoemission spectroscopy (ARPES) and first-principles calculations we find that the spectrum of FeSb$_2$ consists of two bands near the Fermi energy: the nondispersive strongly renormalized $\alpha$-band, and the hole-like $\beta$-band that intersects the first one at $\Gamma$ and Y points of the Brillouin zone. 
Our study reveals the presence of sizable correlations, predominantly among electrons derived from Fe-3d states, and considerable anisotropy in the electronic structure of FeSb$_2$. These key ingredients are of fundamental importance in the description of colossal thermopower in FeSb$_2$.
\end{abstract}

\maketitle

\begin{figure}
\centering
 \includegraphics[scale=1.1]{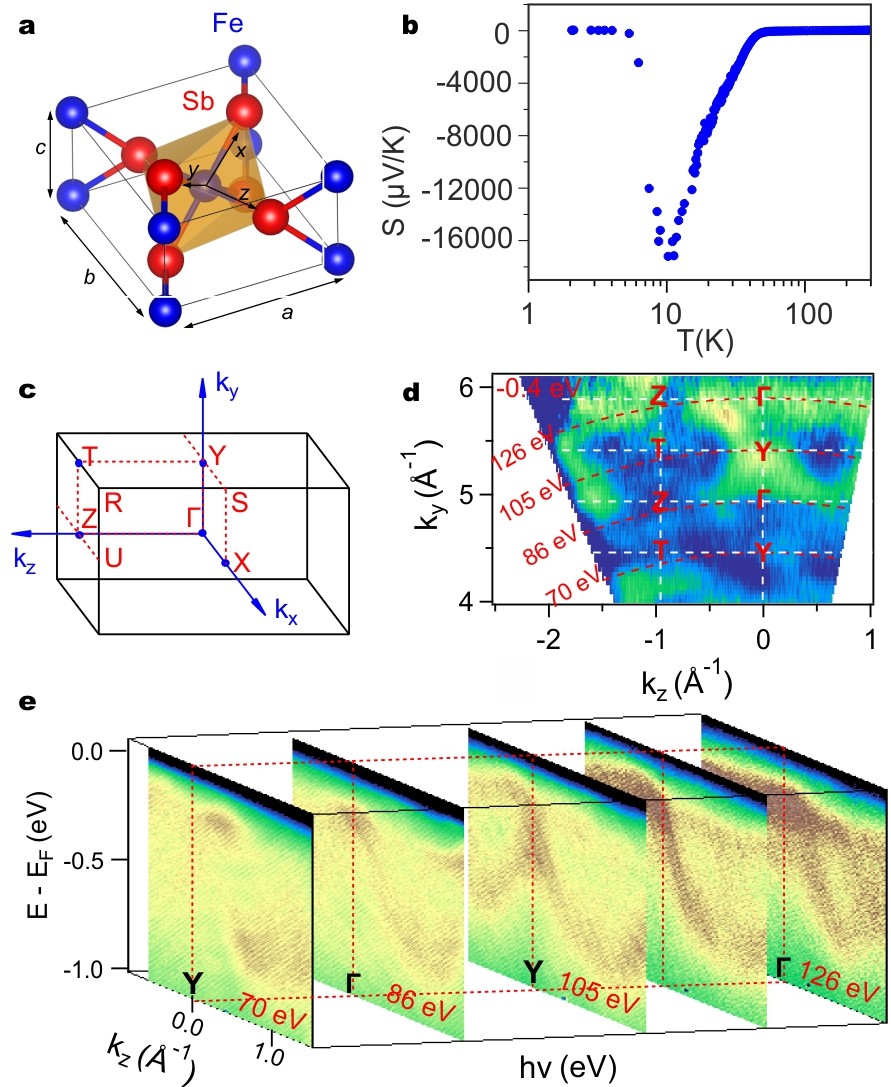}
 \caption{(a) Crystal structure of FeSb$_2$ (space group  $Pnnm$). The unit cell dimensions are a = 5.83 \AA{}, b = 6.51 \AA{}, and c = 3.16  \AA{}. The octahedron formed by the Fe (blue) and the six Sb (red) neighbors atoms is illustrated in orange. (b) Seebeck coefficient ($S$) as a function of temperature ($T$) for thermal gradient along $b$-axis in zero magnetic field for FeSb$_2$. (c) The sketch of the Brillouin zone. (d) Iso-energy surface taken along k$_b$ at -0.4 eV binding energy (BE). (e) 2D ARPES cuts obtained at several photon energies through $\Gamma$-Z direction.}
 \label{fig:Fig1}
\end{figure}

Thermoelectricity in correlated electron materials is of great practical importance as they can exhibit giant thermopower values at low temperatures~\cite{Tomczak3243,PhysRevLett.80.4775,PhysRevLett.87.236603,PhysRevB.82.085104}. This is well exemplified by iron antimonide (FeSb$_2$), which hosts a colossal thermopower variyng  from about (1-2) meV/K to 45 meV/K~\cite{bentien, takahashiNatCom, PhysRevB.86.115121}. The orthorhombic FeSb$_2$ (Fig.~\ref{fig:Fig1}(a)) is a narrow-gap semiconductor that exhibits insulator-to-metal transitions~\cite{PhysRevB.67.155205,Perucchi2006}, unusual magnetic properties~\cite{PhysRevB.72.045103}, and enhanced quasiparticle masses upon chemical substitution~\cite{PhysRevB.74.205105}. Its colossal thermopower is far above the Mott value of the order of $k_{B}/e = 0.08$ mV/K, which is expected and usually recorded in metals or semiconductors~\cite{PhysRevB.82.085104,PhysRev.181.1336}.

It has been proposed that the colossal thermopower of FeSb$_2$ is of electronic origin, mainly in part due to electronic correlations associated with the Fe-3d states~\cite{bentien,PhysRevB.79.153308}.  However, up to date, there is no quantitative theory which incorporates the strong electronic correlations of the Fe-3d electrons while reproducing the value of FeSb$_2$ colossal thermopower. Furthermore, the energy gap $\Delta$ required to explain the experimental thermopower stemming from multiband electronic correlations in the absence of the phonon-mediated vertex corrections is of about 450 meV~\cite{PhysRevB.82.085104}. This value is too large in comparison with values of $\Delta$ between 30 and 70 meV obtained experimentally~\cite{PhysRevB.67.155205,PhysRevB.82.245205,Homes2018}. More recently, phonon-drag mechanisms associated with defect-induced in-gap states have been proposed as the origin of the anomalous thermopower~\cite{PhysRevLett.114.236603}.
Within this picture, the thermopower enhancement up to about 16 mV/K was explained~\cite{PhysRevLett.114.236603,takahashiNatCom}. These two very distinct and conflicting scenarios demonstrate the importance of insight into the electronic structure of FeSb$_2$. 
In this Letter, we disclose for the first time the electronic structure of FeSb$_2$ using a combination of ARPES and many-body first-principles calculations. 

The ARPES measurements were performed at the Surface/Interface Spectroscopy (SIS) X09LA beamline of the Swiss Light Source located at the Paul Scherrer Institute in Villigen, Switzerland. We used photon energies from $60$ eV to $130$ eV with both circular and linear photon polarizations. The measured samples show thermopower peak of about 16 meV/K (Fig. 1(b)) and have been cleaved \textit{in situ} along $ac$-plane (Fig.~\ref{fig:Fig1}(a)) at 17 K. 

\begin{figure}
\centering
 \includegraphics[scale=0.45]{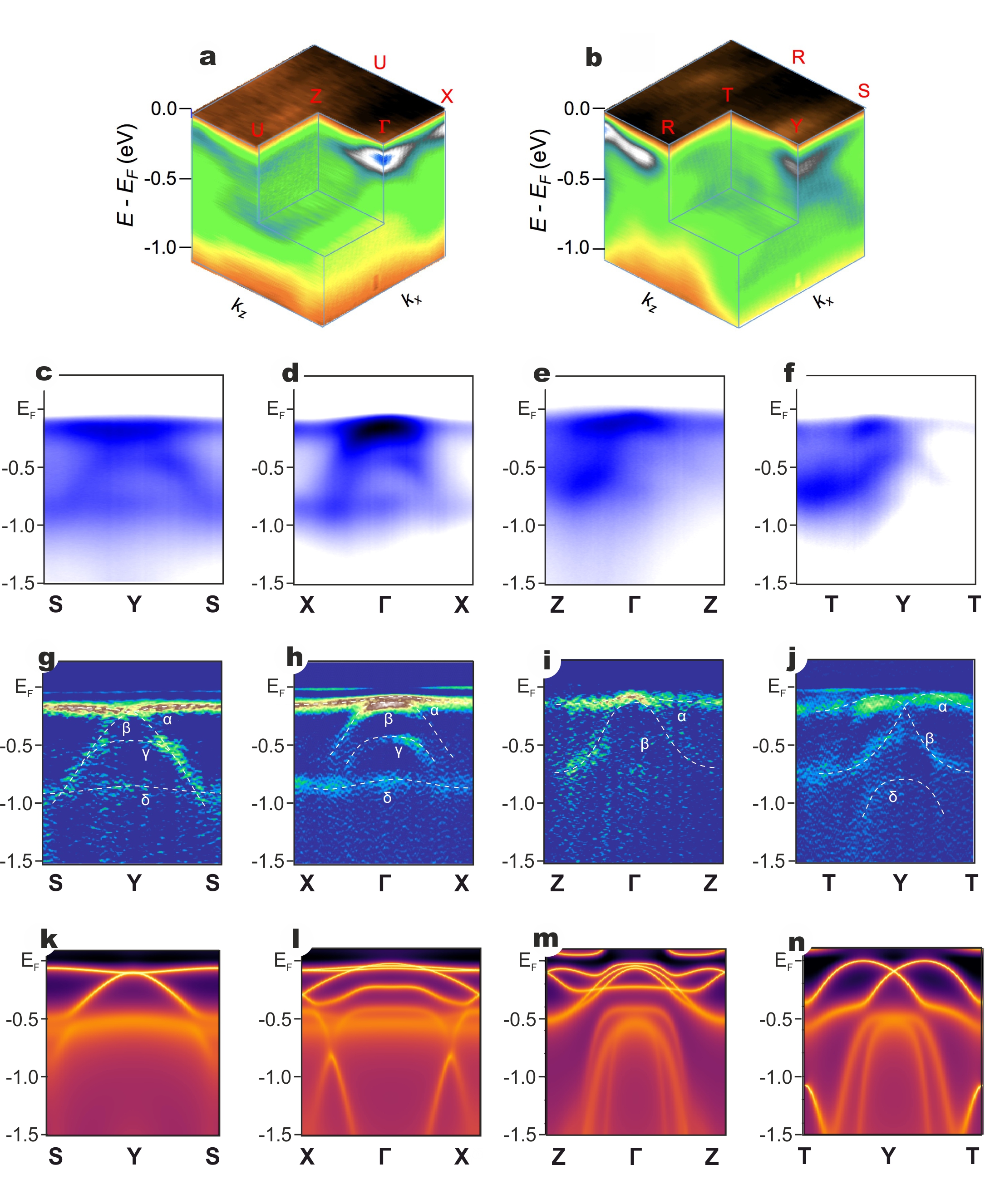}
 \caption{ (a,b)  Full ARPES 3D mapping in k-space. (c-f) ARPES spectra at different high-symmetric cuts in BZ. (g-j) The spectra visualized using the curvature methods.(k-n) LQSGW+DMFT calculated spectral functions at 50 K along the same high-symmetry directions as ARPES data.}
 \label{fig:Fig2}
\end{figure}

Fig.~\ref{fig:Fig1}(d) presents the constant energy cut as a function of photon energy near 0.4 eV BE along k$_b$ crystallographic direction. The corresponding cuts along high-symmetry directions (Fig.~\ref{fig:Fig1}(c)) taken at different photon energies are presented in Fig.~\ref{fig:Fig1}(e). 
Figs.~\ref{fig:Fig2}(a-b) and (c-f) display the experimental ARPES 3D mapping in k-space and cuts along the high-symmetry directions, respectively. Further, using the curvature method the bands are enhanced (Figs.~\ref{fig:Fig2}(g-j)) and compared with the calculations. The experimentally obtained band structure of FeSb$_2$ is composed of two bands near the Fermi energy ($E_{F}$): a light hole-like band (named as $\beta$-band), and a heavy, weakly dispersing band (named as $\alpha$-band)(Fig.~\ref{fig:Fig2}(g-j)). Both do not cross $E_F$ at low temperatures and, consequently, a gap of around $0.05$ eV is formed. These two bands intersect at the $\Gamma$-point in $\Gamma$XUZ plane and near the Y-point in YSRT plane.  The third hole-like $\gamma$-band emerges at $\Gamma$ and Y points (Fig.~\ref{fig:Fig2}(g)) at the same binding energy (BE) (around -0.5 eV) and has the weak dispersion along k$_b$ direction. This $\gamma$-band intersects with the other weakly-dispersing $\delta$-band at $\sim$-0.8 eV of BE. 

In order to get insight into the electronic structure of FeSb$_2$, ARPES data are directly compared with band structures obtained within density functional theory (DFT), linearized quasiparticle self-consistent GW (LQSGW) methods, and LQSGW+DFMT-based spectral functions evaluated at 50 K~\cite{supple}. First, the DFT(LDA) method fail to describe the semiconducting nature of FeSb$_2$. This can be overcome by using the modified Becke-Johnson (mBJ) exchange-correlation potential~\cite{beckejohnson}, which is better in describing the band gaps of semiconductors than LDA.  For FeSb$_2$ the DFT(LDA-mBJ) method overestimats band gap of 0.20 eV.  Further, one can notice that the occupied part of the band structures (bands below zero) (Supplemental material Figs. 1(a-d)) calculated within DFT (LDA) agrees better with the ARPES data than the one obtained with DFT(LDA-mBJ). For instance, the DFT(LDA-mBJ) band structure along Y-S (Supplemental material Fig. 1(a)) shows a dispersing band near $E_F$ and an almost flat 
band (reminiscent of  $\alpha$-band) at -0.88 eV at Y point, in disagreement with ARPES. Although the mBJ exchange-correlation potential allows the correct description of the insulating nature of FeSb$_2$, the electronic correlations taken into account within the mBJ approximation cannot describe the experimental electronic structure of FeSb$_2$. 

\begin{figure}
\centering
 \includegraphics[scale=0.23]{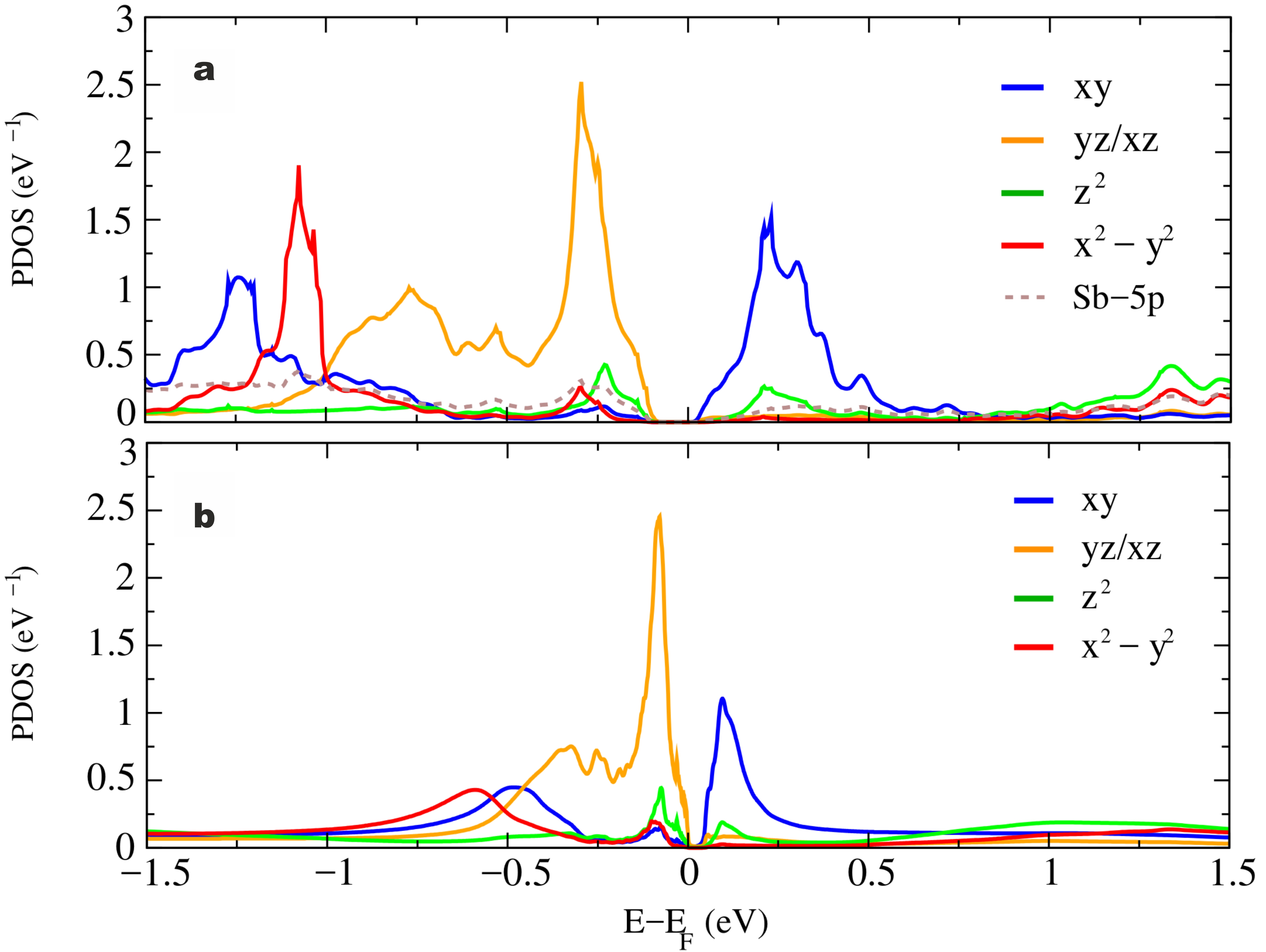}
 \caption{(a) LQSGW and (b) LQSGW+DMFT based projected density of states . The projections to Fe-3d$_{xy}$, Fe-3d$_{yz}$/d$_{xz}$, Fe-3d$_{z^2}$, and Fe-3d$_{x^{2}-y^{2}}$ are shown in blue, orange, green, and red. In (b) the projection to Sb-5p states is shown in dashed brown lines.  These projected density of states were obtained using the local axis shown in Fig.~\ref{fig:Fig1}(a).}
 \label{fig:Fig3}
\end{figure}

Within the LQSGW approximation~\cite{KUTEPOV2017407,PhysRevB.85.155129} we find that FeSb$_2$ has a gap of  $\sim$ 160 meV with occupied part of the spectrum in better agreement with the ARPES data, even though, the corresponding $\alpha$ and $\beta$-bands do not cross exactly at $\Gamma$ and Y points (Supplemental material Figs.1(e-h)). Also, LQSGW calculation predicts the additional bands around -0.5 eV and -1 eV (at $\Gamma$ point) which are reminiscent of $\gamma$ and $\delta$-bands observed in ARPES, as can be seen in Figs.~\ref{fig:Fig2} (g) and (h). The orbital character of each band can be obtained by calculating the projected density of states on the Fe-3d and Sb-5p states. Fig.~\ref{fig:Fig3}(a) presents the LQSGW based projected density of states, where we used the local axis shown in Fig.~\ref{fig:Fig1}(a). As can be noticed, the top of the valence band is mainly due to Fe-3d$_{yz}$/d$_{xz}$ states while the bottom of the conduction band is mainly due to Fe-3d$_{xy}$ states.  The band around -1.
25 eV observed in band structure calculated by the LQSGW (Supplemental material Fig. 1(e) and (f)) is due to Fe-3d$_{xy}$ states, indicating that this band is a bonding state-like while the conduction states are antibonding like. In addition, the band around -1 eV is mainly from the Fe-3d$_{x^{2}-y^{2}}$ states. Finally, it is noteworthy that the Sb-5p states have a minor contribution to the bands around the Fermi energy.

In order to achieve a better agreement between our ARPES data and the calculated electronic structure, the electronic correlations beyond a many-body perturbation theory have to be included. As shown in Figs.~\ref{fig:Fig2}(k-n), the LQSGW+DMFT calculations predict spectral functions which are in better agreement with the experimental band structure. This provides unambiguous evidence of considerable band renormalization in FeSb$_2$. Besides the gap of around 70 meV, the low-energy part of the LQSGW+DMFT spectral functions (Figs.~\ref{fig:Fig2}(i-l)) shows the presence of a nondispersive band and a weakly dispersing band near $E_F$.  Our LQSGW+DMFT calculations also suggest that these bands hybridize near $\Gamma$ and Y points (Figs.~\ref{fig:Fig2}(k) and (l)).  Furthermore, the calculation finds a more dispersive behavior of the $\alpha$-band along the Y-T direction in comparison with the almost flat behavior of the same band along S-Y and $\Gamma$-X, which in good agreement with the experiment (Figs.~\ref{
fig:Fig2}(g-j)). 
Thus, our ARPES data and calculated spectral functions indicate a low-dimensional behavior of the electronic structure of FeSb$_2$. The corresponding anisotropy was recently found in optical conductivity measurements of FeSb$_2$ single crystals, which is in good agreement with one-dimensional semiconducting behavior of the optical properties along the $b$ axis of the $Pnnm$ unit cell~\cite{Homes2018}. As pointed out by early investigations~\cite{PhysRevB.47.16631,KimJAP} low-dimensional behavior favors the appearance of high thermopower.

Our calculations do not capture the presence of high-energy $\gamma$ and $\delta$- bands due to the incoherent nature of the electronic states below -0.4 eV BE, which are reminiscent of Hubbard-like bands. From the projected density of states shown in Fig.~\ref{fig:Fig3}(b) we observe the same orbital character of the valence as conduction band as we found within the LQSGW band structure, i.e. valence band of Fe-3d$_{yz}$/d$_{xz}$ character while conduction band is mainly due to Fe-3d$_{xy}$ states. We also notice that the $\alpha$-band gives rise to a sharp peak near $E_{F}$ which is a common feature among the materials with high thermopowers~\cite{Mahan7436}.
Another important feature is that the weakly dispersing $\alpha$-band is more gaped at the Y-point then at $\Gamma$-point. This can be due to the facts that the strongly dispersing $\beta$-band approaches the Fermi energy at the $\Gamma$-point, which consequently “push-ups” the $\alpha$-band, as seen in Figs.~\ref{fig:Fig2}(h) and (i). In addition, as we show in Figs.~\ref{fig:Fig2} (a) and (b), the valence band maximum occurs at the $\Gamma$ point, in contrast to band structures calculated using LQSGW where the top of the valence band is located at R point (Supplemental material Figs.5(a) and (b)). Yet, the calculated spectrum using LQSGW+DMFT method, in turn, agrees with our experimental findings at this point, as shown in Fig.5(c) and (d) in the supplemental material.
Hence, the calculations outlined in this Letter demonstrate that a proper theoretical description of the electronic structure of FeSb$_2$ requires a treatment beyond a perturbative approximation. Indeed, our LQSGW+DMFT calculations predict sizable electron-electron correlations among the Fe-3d states of FeSb$_2$, with $Z_{DMFT} \sim 0.5$, which cannot be captured within the LQSGW method.

\begin{figure}
\centering
 \includegraphics[scale=0.45]{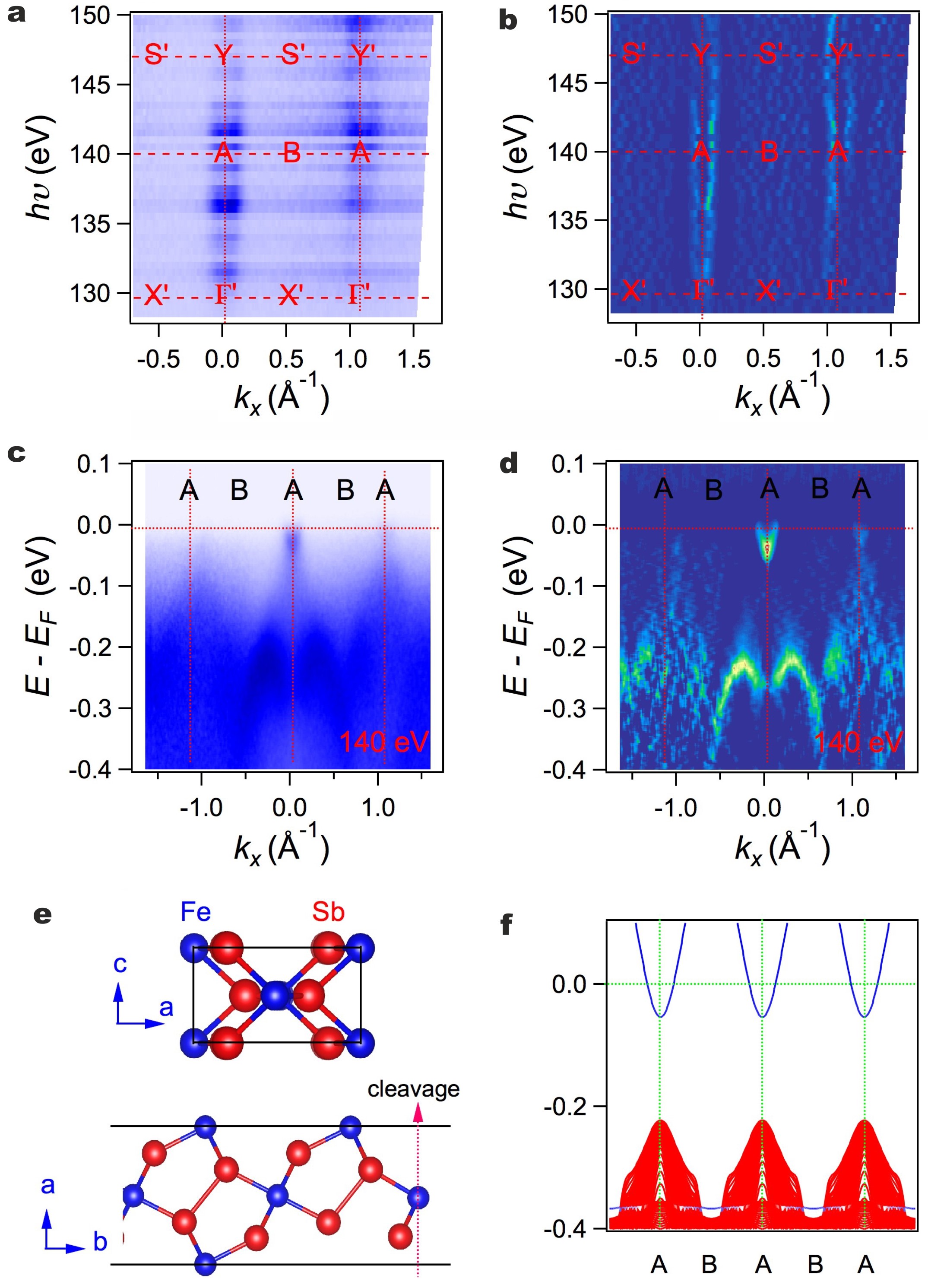}
 \caption{ (a) The raw iso-energy surface taken along k$_b$ at Fermi level and (b) the same data visualized with the curvature method. (c) The raw ARPES spectra taken along A-B cut (see Fig.~\ref{fig:Fig5}(a,d)) and (d) visualized with the curvature method. (e) Fe-terminated surface. Iron and antimony atoms are represented by blue and red spheres, respectively; (f) FeSb$_2$ Slab band structure calculated within tight-binding approximation parametrized from our DFT calculations. The surface bands are shown in blue whereas the bulk bands are shown in red.}
 \label{fig:Fig5}
\end{figure}

Although we clearly observe the valence band maximum at $\sim$-50 meV BE, the sample surprisingly showed no charging effect during ARPES experiments. This evidence shows that the measured system is, in fact, metallic suggesting a presence of some states crossing the Fermi level. Indeed we found the two-dimensional (2D) band, at $ac$-plain responsible for the surface conductivity. Figures~\ref{fig:Fig5}(a) and (d) depict spectral intensity distribution at Fermi energy pointing out the 2D-, and, therefore,surface-character of this state. The cut is taken at 140 eV photon energy (Figs.~\ref{fig:Fig5}(b) and (e)) clearly shows the electron-like pocket at A-B-A direction, which is absent in the bulk spectral function calculated along the same direction (Supplemental material Fig.6).

In order to disentangle the origin of the surface states observed by ARPES we have performed additional DFT and tight-binding calculations for a FeSb$_2$ slab terminated by Fe atoms (Fig.~\ref{fig:Fig5}(e)). Calculated band structure (Fig.~\ref{fig:Fig5} (f)) reveals the presence of the electron-like surface band (blue line). The electron pocket derived from this band at the A point of BZ is an agreement with experimental data. Importantly, this surface band is mainly formed by Fe-3d states at the surface, while the Sb-5p states give a minor contribution to this band (Supplemental material Fig.4).

In summary, we performed the detailed experimental and theoretical investigation of the FeSb$_2$ electronic structure. Obtained ARPES data reveal that the electronic structure of FeSb$_2$ near the Fermi energy consists of two bands: the strongly renormalized and weakly dispersing $\alpha$-band and the $\beta$-band which is more dispersive and behaves like a light hole-like band.
The LQSGW+DMFT calculations agree well with the experimentally depicted electronic structure providing the clear evidence of sizable electronic correlations ($Z_{DMFT} \sim ~ 0.5$) in the Fe-3d states. Importantly, our ARPES data and calculated spectral functions demonstrate the anisotropy of the electronic structure causing the low-dimensional behavior of FeSb$_2$, which may favor the appearance of anisotropic thermopower response.  The combination of advanced experimental and theoretical studies establishes the fundamental ingredients in understanding FeSb$_2$ colossal thermopower phenomenon. 

\textit{Acknowledgments.--} This work was supported by the U.S. Department of Energy, Office of Science, Basic Energy Sciences as a part of the Computational Materials Science Program. This research used resources of the National Energy Research Scientific Computing Center (NERSC), a U.S. Department of Energy Office of Science User Facility operated under Contract No. DE-AC02-05CH11231.
ARPES Experiments were conducted at the Surface/Interface Spectroscopy (SIS) beam line of the Swiss Light Source at the Paul Scherrer Institut in Villigen, Switzerland. The authors thank the technical staff at the SIS beam line for their support.

\newpage 
\newpage 

\section{Supplemental Material}

\subsection{First principles methods}

\subsubsection{FeSb$_2$ bulk}
Our DFT calculations were performed within the local-density approximation(LDA)~\cite{kohn} and modified Becke-Johnson (LDA-mBJ) exchange-correlation potential~\cite{beckejohnson}, as implemented in Wien2k~\cite{Wien2k}. The linearized quasiparticle self-consistent GW (LQSGW) calculations were performed using the FlapwMBPT code~\cite{PhysRevB.80.041103,PhysRevB.85.155129,KUTEPOV2017407} where the Muffin-tin radii in Bohr radius are $2.6$ and $2.2$ for Fe and Sb, respectively. In the calculation of polarizability and self-energy, unoccupied states with an energy up to 200 eV from the Fermi energy were taken into account.  
In our LQSGW+DMFT calculations performed within COMSUITE~\cite{schoi,comsuite}, projectors to the correlated Fe-3d orbitals were constructed using Fe-3d and Sb-5p based maximally-localized Wannier functions~\cite{RevModPhys.84.1419}. These projectors span the electronic states in the energy window of $E_{F} \pm 8$ eV in a $15 \times 15 \times 30$ k-grid. It is important to mention that by using this large energy window we construct very localized Fe-3d orbitals.
Next, we evaluate the local self-energy associated with the Fe-3d orbitals within dynamical mean field theory (DMFT)~\cite{DMFT} using static $U_d$ and $J_H$. These two quantities are evaluated by using a modification of the constrained random phase approximation~\cite{schoi,crpa}, which avoids the screening from the correlated as well as hybridized bands. In particular, by using Slater's integrals~\cite{PhysRevB.37.10674, PhysRevB.82.045105} we obtain $U_{d} = 4.8$ eV and $J_{H} = 1.0$ eV. Finally, we mention that the Feynman diagrams included in both LQSGW and DMFT (double counting) are the local Hartree and local GW diagrams. They are computed using the local projection of the LQSGW's Green's function and the local Coulomb matrix constructed by $U_d$ and $J_H$. 

\subsubsection{FeSb$_2$ surface}

Our DFT calculations for the FeSb$_2$ surface were performed within the  Perdew-Burke-Ernzehof generalized gradient approximation (PBE-GGA)~\cite{pbe} using the Quantum Espresso suite~\cite{qespresso}. In our calculations a plane wave basis-set with energy cutoff of 30 Ry was employed. The atomic positions were relaxed until the total forces on each atom were smaller than 10$^{-3}$ a.u.. Further, in our calculations the electron-ion interactions were described through ultrasoft pseudopotentials~\cite{PhysRevB.41.7892}.

\newpage 
\subsection{DFT and LQSGW  band structures}

\begin{figure*}[h]
\centering
 \includegraphics[scale=0.5]{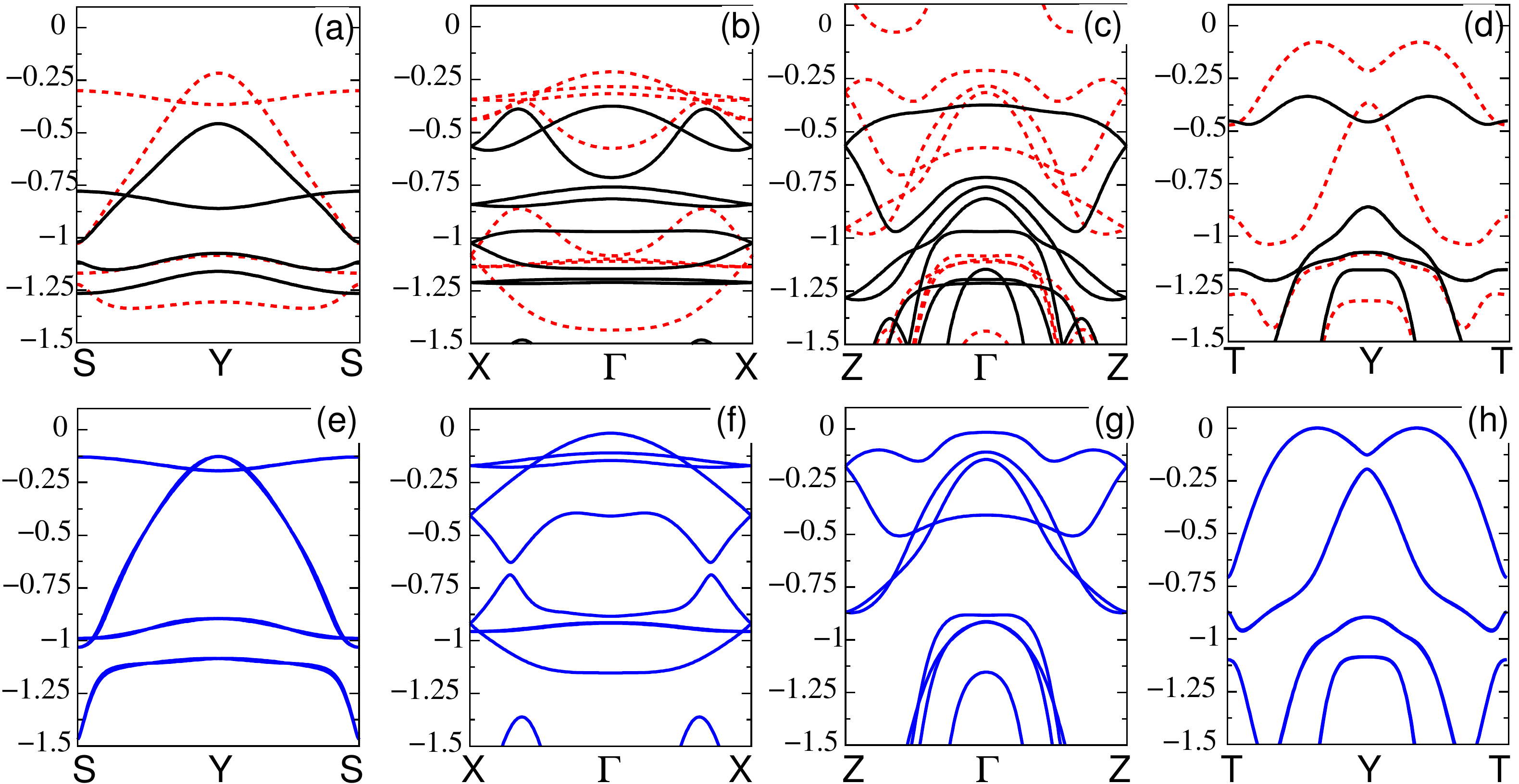}
 \caption{(a-d) DFT band structures within LDA (dashed red lines) and LDA-mBJ exchange correlation potential (black lines). In (e-h) we show the LQSGW band structures (blue lines).}
 \label{fig:FigS1}
\end{figure*}

Our calculated DFT(LDA), DFT(LDA-mBJ) and LQSGW band structures are shown Fig.~\ref{fig:FigS1}.  These figures display the electronic band structures along Y-S, $\Gamma$-X, $\Gamma$-Z, and Y-T high symmetry directions of Brillouin zone (see Fig.1(c) in the main text). As we discussed in the main text our DFT(LDA) fails to capture the insulating nature of FeSb$_2$ whereas DFT(LDA-mBJ) calculations predict an overestimated band gap of 0.20 eV. The value of the band gap is in better agreement with experiments when LQSGW is used, where we find a band gap of $\sim$ 160 meV. More important, the occupied bands are in better agreement with the ARPES data as shown in the main text.

\begin{figure*}[h]
\centering
 \includegraphics[scale=0.75]{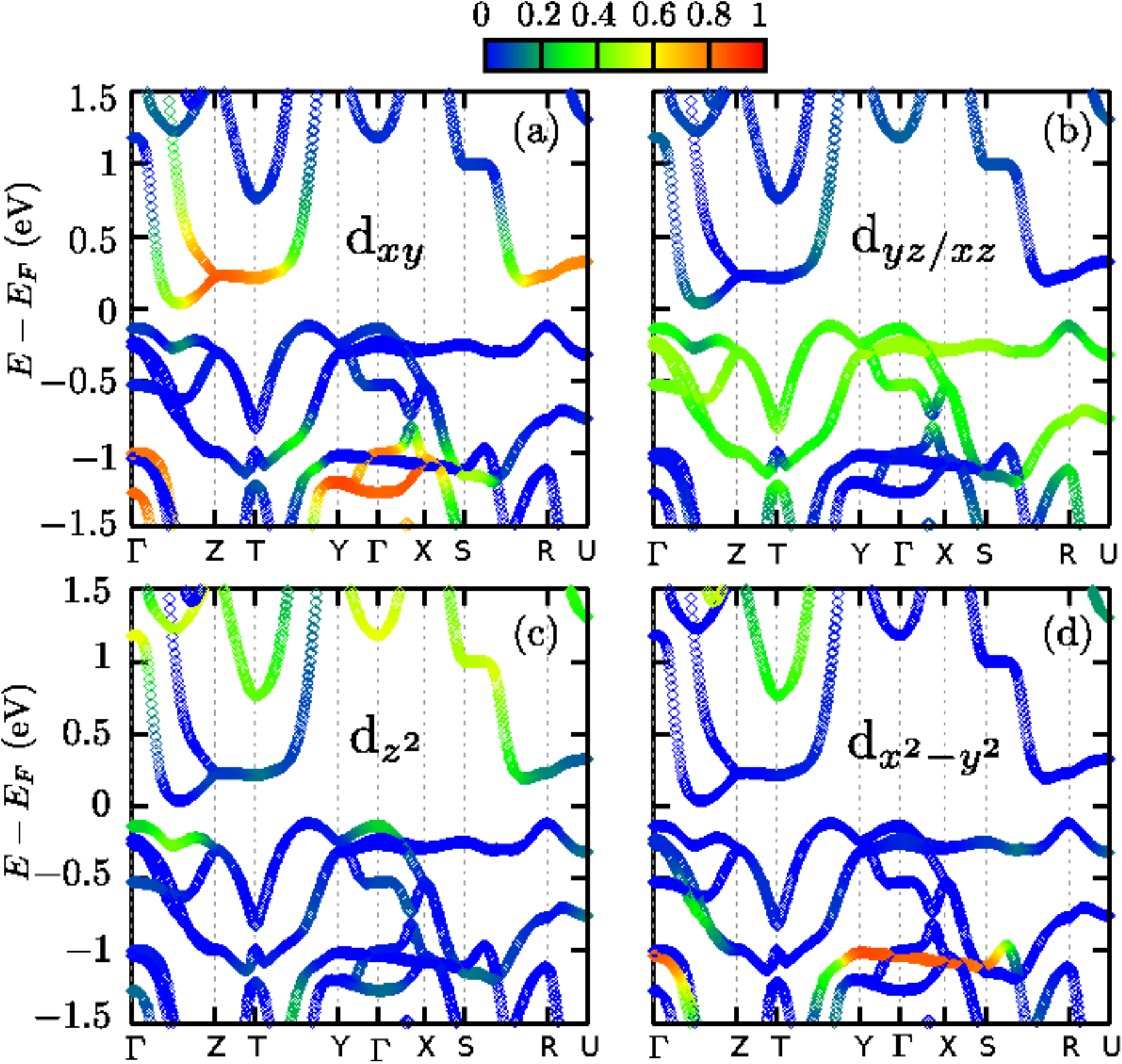}
 \caption{Fe-3d orbital resolved LQSGW band structure of FeSb$_2$. We present the contribution to the iron d$_{xy}$, d$_{yz}$/d$_{xz}$, d$_{z^{2}}$, and d$_{x^{2}-y^{2}}$ in (a), (b), (c), and (d), respectively.}
 \label{fig:FigS2}
\end{figure*}

\begin{figure*}[h]
\centering
 \includegraphics[scale=0.4]{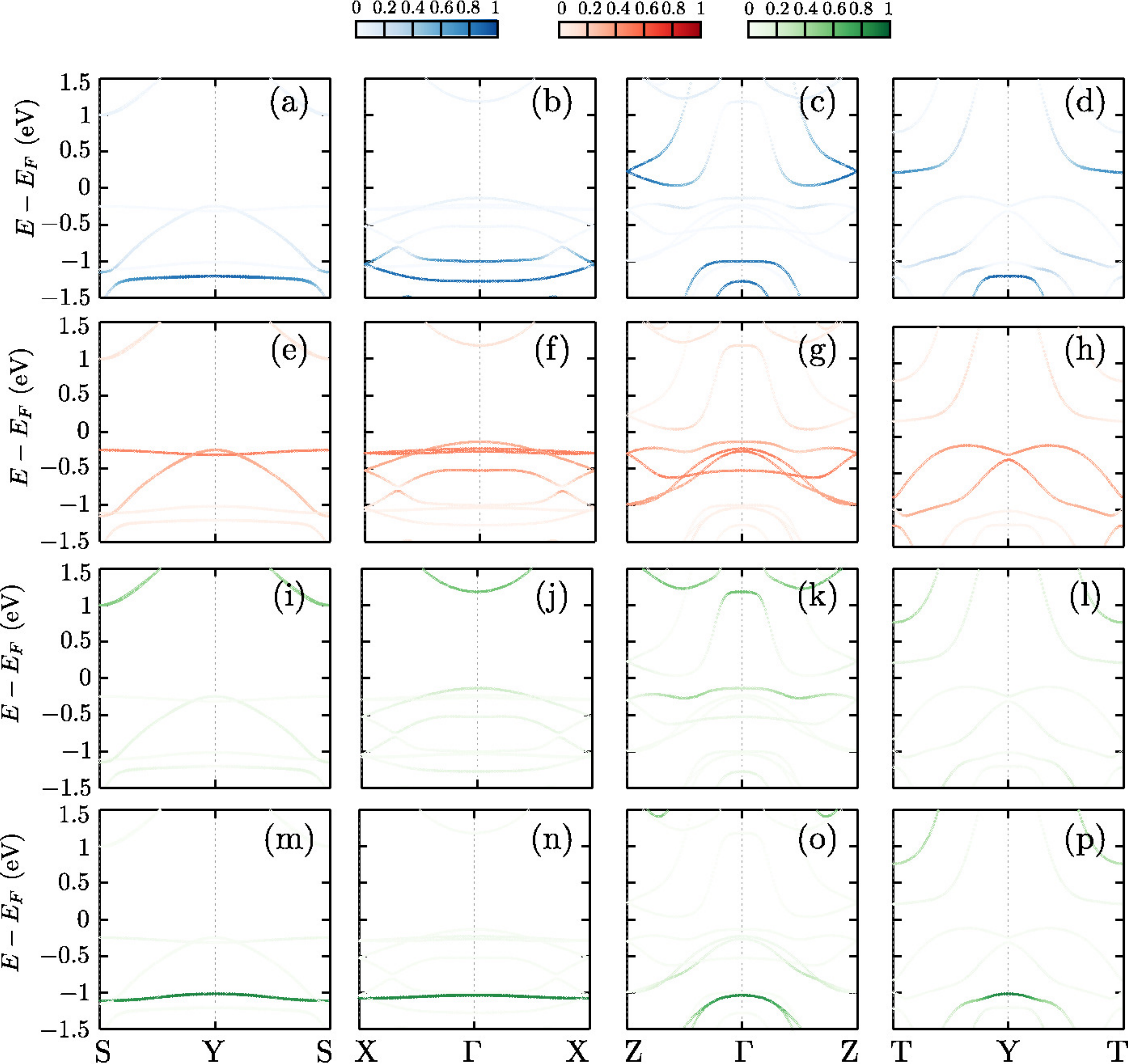}
 \caption{Fe-3d orbital resolved LQSGW band structure along the same high-symmetry directions in BZ as ARPES data.
 Projections to Fe-d$_{xy}$ (a-d), Fe-d$_{yz}$/d$_{xz}$ (e-h), Fe-d$_{z^2}$ (i-l), and Fe-d$_{x^{2}-y^{2}}$,are shown in blue, red, and green, respectively.}
 \label{fig:FigS3}
\end{figure*}

The orbital resolved LQSGW band structure are shown in Fig.~\ref{fig:FigS2} and Fig.~\ref{fig:FigS3}. From these figures one can observe that the valence band is mainly composed by Fe-d$_{yz}$/d$_{xz}$ states whereas the conduction band is mainly of Fe-d$_{xy}$ character, in agreement with the LQSGW projected density of states (see Fig.3(a) in the main text).

\newpage

\subsection{DFT calculations for FeSb$_2$ surface}

\begin{figure*}[h]
\centering
 \includegraphics[scale=0.7]{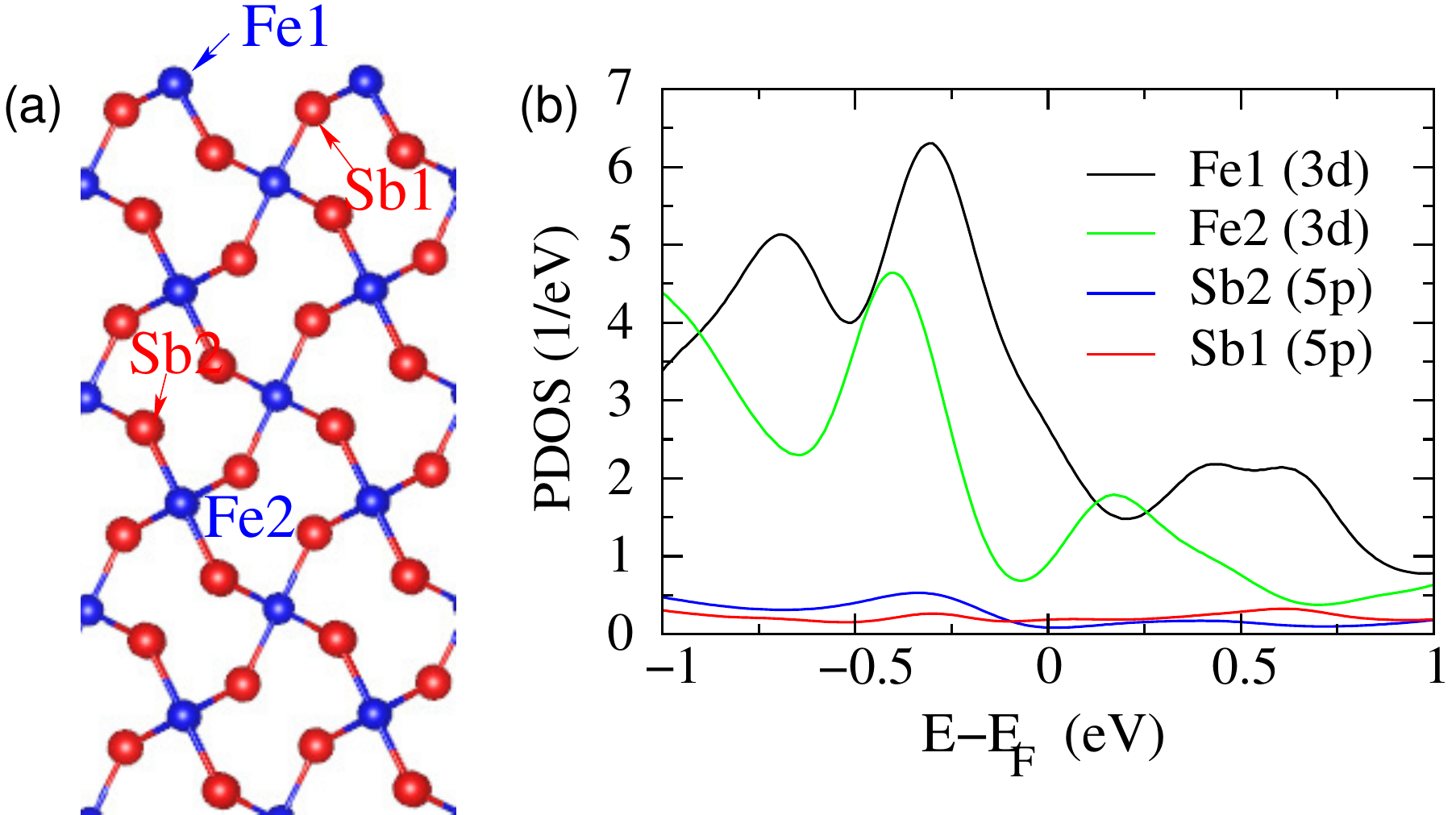}
 \caption{(a) Theoretical slab model for the FeSb$_2$ surface which is terminated in Fe (blue spheres). (b) Projected density of states on the Fe1, Fe2, Sb1, and Sb2 sites illustrated in (a).}
 \label{fig:FigS5}
\end{figure*}

In our investigation on the FeSb$_2$ surface we initially employed DFT calculations to relax the atomic positions of a slab generated through a $ac$-cleaveage plane. In Fig.~\ref{fig:FigS5}(a) we illustrate the resulting iron terminated surface.
As pointed out in the main text, the metallic surface states are mainly due to the Fe-3d states of Fe1 site, as shown in the projected density of states in Fig.~\ref{fig:FigS5}(b). 

\subsection{LQSGW+DMFT spectral function}

\begin{figure*}[h]
    \centering
    \includegraphics[scale=0.6]{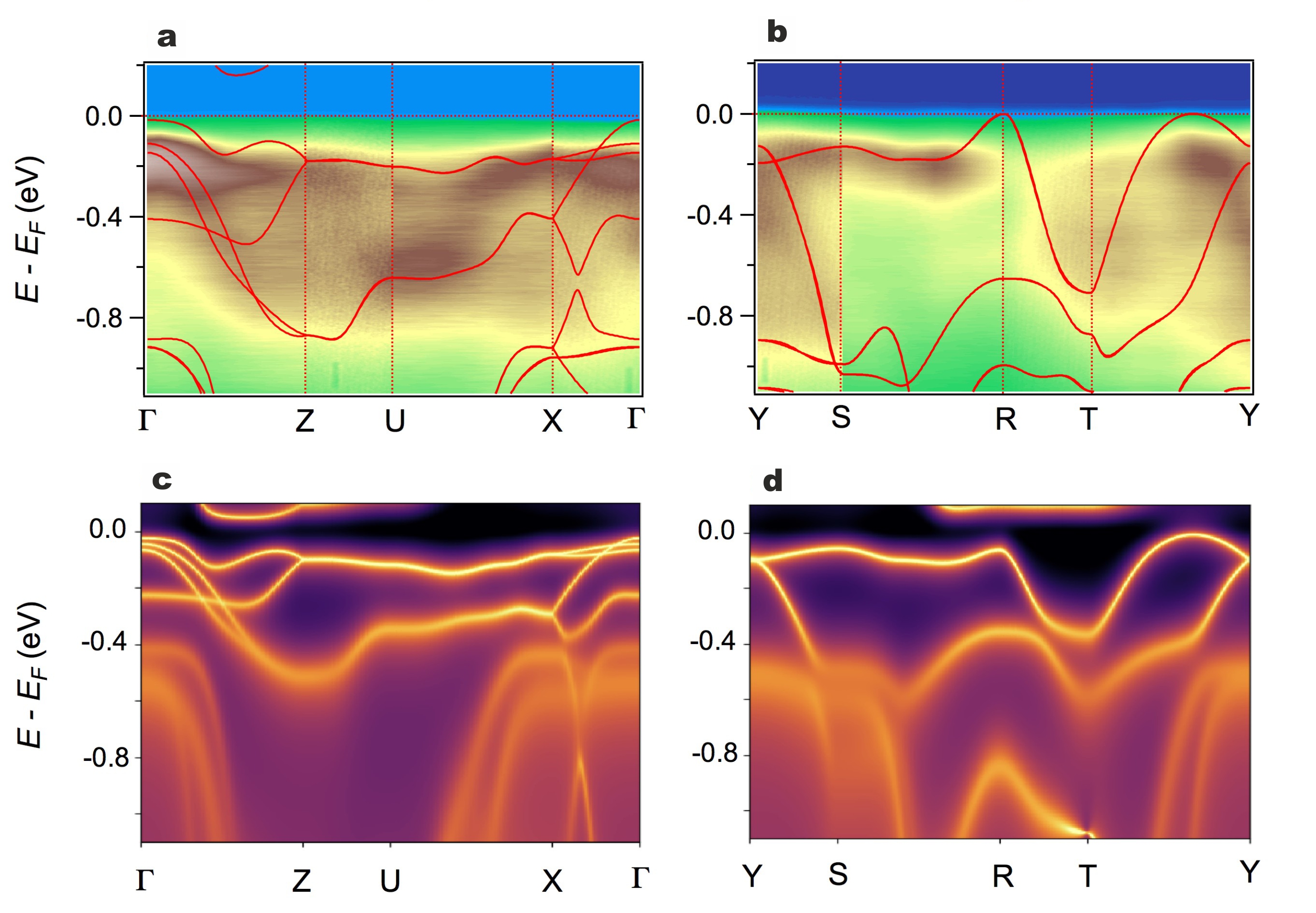}
    \caption{ARPES derived spectra together with calculated LQSGW band structure of FeSb$_2$ along $\Gamma$-Z-U-X-$\Gamma$ (a) and Y-S-R-T-Y (b). In (c) and (d) we show the LQSGW+DMFT spectral functions at 50 K along the same directions.}
    \label{fig:FigS6}
\end{figure*}

As discussed in the main text our ARPES data indicates the maximum of the valence band takes place at the $\Gamma$ point. As shown in Figure~\ref{fig:FigS6}(a) and (b) our LQSGW band structure predicts that the maximum takes place at the R, in contrast with our ARPES data. On the other hand, one can observe that the calculated LQSGW+DMFT spectral functions predict the valence band maximum at the $\Gamma$ point (Fig.~\ref{fig:FigS6}(c) and (d)) in better agreement with our experimental findings.

\begin{figure}[h]
\centering
 \includegraphics[scale=0.4]{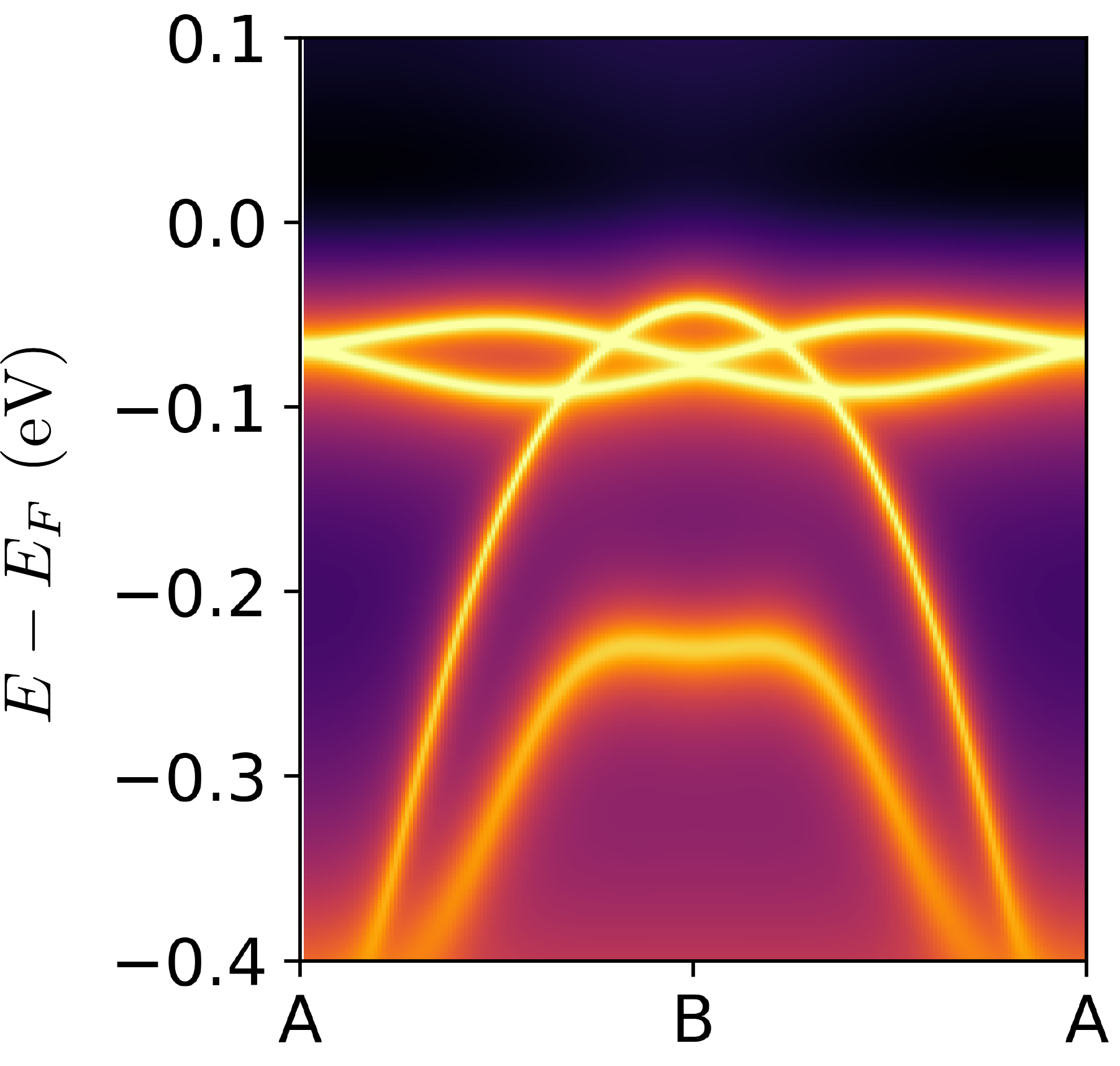}
 \caption{LQSGW+DMFT spectral function of FeSb$_2$ bulk along A-B-A at 50 K.}
 \label{fig:FigS4}
\end{figure}

Furthermore, in Fig.~\ref{fig:FigS4} we show the LQSGW+DMFT spectral function calculated along A-B-A direction for the FeSb$_2$ bulk. As pointed out in the main text, one can see there is no electron pocket at the A point.

\end{document}